\documentclass[a4paper,12pt]{article}
\usepackage{amsbsy}
\usepackage[left=2cm,top=1.75cm,right=1cm,bottom=1.75cm]{geometry}
\usepackage{graphics}
\usepackage{graphicx}
\usepackage{epsfig}
\usepackage{amsmath}
\title{ Effects of  spatially inhomogeneous atomic interactions on Bose-Einstein condensates in optical lattices}
\author{{\small \bf Golam Ali Sekh\footnote{Email: skgolamali@gmail.com}}\\
{\small \it  Dipartimento di Fisica "E. R. Caianiello", via ponte don Melillo I-84084, Fisciano
(SA), Italy}}
\date{}
\begin{document}
\maketitle
\begin{abstract}
An interplay of optical lattices and nonlinear impurities in controlling the dynamics of Bose-Einstein condensate  bright solitons is investigated using  effective potential approach. The ability of pushing the solitons into or away from the impurity region by changing both lattice and impurity parameters is suggested. A possibility for the existence of stable fundamental gap solitons, which appear to satisfy  an inverted Vakhitov-Kolokolov criterion, is examined.
\end{abstract}
\section{Introduction}
The first experimental observation of Bose-Einstein condensates (BECs) in 1995 \cite{1} is one of the great achievements of modern science. This milestone achievement has been motivated many  theoretical and experimental works \cite{2}  including creation and propagation of matter-wave solitons  \cite{3}. Solitons  in a BEC are traditionally produced by changing the sign of atom-atom scattering length ($a_s$) from positive to negative values or vice versa with  the help of magnetic \cite{4} or optical \cite{5} Feshbach resonances (FRs). These FRs either individually or a combination \cite{6} of them offer an additional flexibility to obtain spatial variation of the scattering length and thereby introduces spatially inhomogeneous nonlinear interaction in the condensate. BECs  with such nonlinear interactions have become  systems of considerable current interest since it is expected that these systems may help to understand the dynamics of solitons emission which  will be useful to construct atom laser \cite{7,8}. In view of this, some theoretical works have been done on the existence, stability and localization of matter-wave solions in  single -component \cite{9,10,11} as well as  two-component \cite{12,13,14} BECs with spatially spatial inhomogeneous atomic interaction.
\par
A BEC can also be loaded in periodic potentials. These potentials, often called optical lattices (OLs), are  generally created by interfering two counter propagating laser beams \cite{15}. The geometry and depth of the OL can be  completely controlled \cite{16} to create a large variety optical potentials. Studies in the dynamics of BEC solitons in periodic  potentials
with different spatially inhomogeneous atomic interaction have received a great deal of attention. Depending on the spatial variation of the atomic interactions, this system shows some new and interesting phenomena. For example, a localized state of BECs loaded in OLs can undergo  delocalized transitions \cite{17} when the atomic interaction is made to vary periodically in space. This inhomogeneous interaction helps  a BEC in OLs to support gap solitons in the upper half of the forbidden gap even when $a_0$, the background atomic scattering,  vanishes \cite{18}.  A special feature of a bound state in a  BEC with localized inhomogeneous interaction is that it can contain any arbitrary number of solitons \cite{8}. Moreover, the potential arising from the spatially inhomogeneous atomic interaction may help a BEC to support  bright as well as dark soliton solutions \cite{19}. Recently, Sakaguchi et al investigated the existence, stability and mobility of solitons in single-component BECs for spatial periodically modulated nonlinearity \cite{20}. For two-component BECs studies on the stability properties of solitons in OLs with spatially periodic interatomic and inter-species interactions have also been made in \cite{21}. The object of the present paper is to study the dynamics of matter-wave  solitons in  OLs with spatially localized  (Gaussian) nonlinear interactions within the framework of variational approach. In particular, we  work with a cigar-shaped  BEC and make use of a potential model to investigate effects of the optical lattice wave number as well as  width of the Gaussian nonlinearity on  localization of the BEC soliton. In addition, we examine that the stable
fundamnetal gap soliton supported by the system satisfy an  anti Vakhitov-Kolokolov criterion.
\par
The mean-field dynamics of a cigar-shaped BEC is satisfactorily described by means of a quasi-one-dimensional (Q1D) Gross-Pitaevskii (GP) equation \cite{22}. The GP equation for a cigar-shaped BEC in optical lattices is given by
\begin{equation}
i \frac{\partial \phi}{\partial \tau}=-\frac{1}{2}\frac{\partial^2 \phi}{\partial s^2}+ V_0 \cos(2 k_L s)\phi +g_{1D}|\phi|^2 \phi,
\end{equation}
where $\phi(=\phi(s,\tau))$ is the order parameter which is normalized to the total number of atoms $N$ in the condensate and  $g_{1D}=2 a_s/a_{\perp}$ \cite{23,24}. The quantities $V_0$ and $ k_L$ stand for the strength and wave number of the OL respectively.  In $(1)$ we have measured length ($s$), time ($\tau$) in the units of $a_\perp $ and $\omega^{-1}_\perp$.  Understandably,  $\omega_\perp$  refers to frequency  of the transverse harmonic confinement that produces the cigar-shaped BEC and $a_\perp=\sqrt{\hbar/m\omega_\perp}$ with $m$, the atomic mass.
\par
As noted above, the mean-field  atomic interaction $g_{1D}$ in $(1)$ can be varied spatially only by changing the s-wave scattering length $a_s$. Experimentally, this is achieved  by using a proper detuned laser light in the vicinity of the FRs.  The dependence of $a_s$ on the spatial coordinate $s$ is given by \cite{5,25}
\begin{equation}
 a(s)=a_0+ \frac{\alpha I(s)}{\delta+\beta I(s)},
\end{equation}
where, $I(s)$ is the intensity of the laser light and $a_0$ stands for the scattering length in the absence of the light. The quantities $\alpha$ and $\beta$ are the constants which depend on the detuning parameter $\delta$ of the laser. Clearly, such variation of $a(s)$ will make $g_{1D}$ in $(1)$ space-dependent.
For a large detuning laser beam having Gaussian intensity variation, $g_{1D}$ can be written as
\begin{equation}
 g_{1D}=\gamma_0+\gamma_1 exp\left[-s^2/2 w^2\right] ,
\end{equation}
where $w$ is the width of the Gaussian profile and,   $\gamma_0$  and $\gamma_1$ are the constants. Equation $(1)$ with $g_{1D}$ in $(3)$ will  provide the governing equation for a cigar-shaped BEC in OLs having spatially localized nonlinear interaction.
\par
In section $2$ we  present a variational formulation using a Gaussian ansatz and derive equations for the parameters of the trail solution with the help of Ritz optimization procedure. In section $3$ we construct a potential model from the newtonian equation  for the center of the trial solution and analyze how the effective potential changes with the variation of $(i)$ the wave number of the OL and $(ii)$ width $(w)$  of the laser beam of Gaussian intensity variation. In both cases, we focus our attention on whether the width of the soliton exceeds the period of the lattice or not.
In section $4$ we  consider the dynamics of the condensates in the effective potential. In particular,  we  study time evolution of the  center and  velocity of the soliton within the framework of a variational formulation and simultaneously perform  a purely numerical calculation to examine stability as well as to verify the result obtained from the variational approach. In section $5$ we envisage a study on the existence and stability of fundamental gap soliton in a BEC with spatially localized atomic interaction.
Finally, in section $6$ we  make some concluding remarks.
\section{Variational formulation}
The initial boundary value problem in $(1)$ can be restated as the variational problem
\begin{equation}
\delta \int {\cal{L}}(\phi,\phi^*,\frac{\partial \phi}{\partial s},\frac{\partial
\phi^*}{\partial s},\frac{\partial
\phi}{\partial \tau},\frac{\partial \phi^*}{\partial \tau})ds \,d\tau=0.
\end{equation}
The Lagrangian density ${\cal{L}}$ of $(1)$ obtained from $(4)$ is given by
\begin{equation}
{\cal L}=\frac{1}{2}i \left(\phi\phi^*_\tau- \phi^*\phi_\tau\right)+\frac{1}{2} |\phi_s|^2+\frac{1}{2} g(s)|\phi|^4+ V(s) |\phi|^2.
\end{equation}
Here $\phi_s=\frac{\partial
\phi}{\partial s}$ and $\phi_\tau=\frac{\partial
\phi}{\partial \tau}$.
The solution of a self-focussing nonlinear Schr\"odinger equation is a $sech$ function \cite{26}. The shape of a Gaussian function is almost identical with a $sech$ function. Therefore, in our variational formulation we adopt  a Gaussian function as a trial solution \cite{27} of $(1)$. To include the effects OL and spatial inhomogeneous nonlinearity we make  parameters of the trial solution time-dependent and thus introduce
\begin{equation}
\phi(s,\tau)=A(\tau) e^{\left[-A(\tau)^2 (s-s_0(\tau))^2/2 \right]}\,e^{\left[i\,\dot{s}_0\left(s-s_0(\tau))\right)+i\, \Phi(\tau)\right]},
\end{equation}
as the variational ansatz for the problem. In $(6)$, $A(\tau)$,  $s_0(\tau)$ and  $\Phi(\tau)$  represent  amplitude, center of mass and phase of the solitonic solution and also act as variational parameters in our formulation. Understandably,  $A(\tau)^{-1}$  and $\dot{s}_0(=ds/d\tau)$ stand for the width and velocity of the soliton.
\par
Now we shall make use of the Ritz optimization procedure \cite{27,28} to obtain equations for the variational parameters.
In this optimization procedure,  the first variation of the variational functional is made to vanish within a set of suitably chosen trial functions. Understandably,  the variational functional for the problem considered here is the so-called effective Lagrangian density. Inserting $(6)$ in $(5)$ and then integrating the resulting equation over $s$ from $-\infty$ to $+\infty$ we get
\begin{eqnarray}
\left\langle {\cal L}\right\rangle =\frac{\sqrt{\pi}}{4}A(\tau)\left(8 s_0(\tau) \ddot{s}_0(\tau)+4
 \dot{\Phi}(\tau)-2 {\dot{s_0}(\tau)}^2+4
 e^{-\frac{k^2_L}{ A(\tau)^2}} V_0 \cos\left(2 k_L s_0(\tau)\right)\right.\nonumber\\\left.+
 {A(\tau)^2}(1+{\gamma_0}{\sqrt{2}})+
\frac{2\gamma_1 A(\tau)^3}{\sqrt{1/(2 w^2)+2 A(\tau)^2}}
e^{-\frac{2 A(\tau)^2 s_0^2}{1+4 w^2 A(\tau)^2}}\right)
\end{eqnarray}
as the effective Lagrangian density.
From the vanishing conditions of the variational derivatives of
$\left\langle {\cal L}\right\rangle$ with respect to  $\Phi$, $A(\tau)$ and ${s}_0$, we obtain the following  equations.
\begin{subequations}
 \begin{eqnarray}
\frac{d}{d\,\tau}\left[\sqrt{{\pi}} \, A(\tau) \right] =0,
\end{eqnarray}
\begin{eqnarray}
\frac{d^2{s}_0}{d\tau^2}=-\frac{1}{3}\frac{d}{d\,s_0}\left(V_0 \cos\left(2 k_L s_0\right) e^{-\frac{k_L^2}{2 A(\tau)^2}}+\frac{\gamma_1 A(\tau)^3}{\sqrt{2/w^2+8 A(\tau)^2}} e^{\frac{-2k_L^2 s_0}{1+4 w^2A(\tau)^2}}\right)
\end{eqnarray}
and
\begin{eqnarray}
&&\hspace{0.5cm}\frac{\sqrt{\pi}}{2}\left[8 s_0(\tau)  4 \ddot{s}_0(\tau)+4 \dot{\Phi}(\tau)-2 \dot{s}_0^2+4V_0 \cos(2k_L s_0) e^{-\frac{2 k_L^2}{A^2}}\left(1+\frac{k_L^2}{A^2} \right)
\right.\nonumber\\
&+&\left.{3 A^2}\left(1+{\sqrt{2}{\gamma_0}} \right)+\frac{8\sqrt{2}\gamma_1 A^3}{\sqrt{1/w^2+8 A^2}} e^{-\frac{2 A^2 s_0^2}{1+4 w^2A^2}}\frac{1+12 A^4 w^4+7A^2 w^2-s_0^2A^2}{1+4 w^2 A^2}\right] =0.
\end{eqnarray}
\end{subequations}
Clearly, these equations show  how the parameters of the lattice and a BEC soliton  are related with each other. It is interesting to note  that the equations in $(8)$  provide a natural basis to study the dynamics of the solitons.  For example, $(8b)$ is a newtonian equation that describes the evolution of the soliton's center in the presence of localized nonlinearity and an OL while $(8a)$ represents an amplitude equation which shows that during evolution number of atoms $N$ in the soliton remains constant since normalization condition of $(6)$ gives $\sqrt{\pi}A(\tau)=N$. On the other hand, $(8c)$ describes the phase evolution of the condensate. In view of this, we shall judiciously make use of $(8)$ to construct an effective potential for the soliton and then study their dynamics for different environment of the effective potential.
\section{Potential model}
Potential model  is one of the simplest analytical models that  may be used to understand  dynamics of optical  as well as BEC solitons \cite{13,14,27,29}. In fact, this model is employed to calculate an effective potential for a moving soliton by drawing  an analogy with the equation of motion of a classical particle. The effective potential of the soliton  obtained from $(8b)$ is given by
\begin{eqnarray}
V_{\rm eff}=\frac{1}{3}\left(V_0 \cos\left(2 k_L s_0\right) e^{-\frac{k_L^2}{2 A(\tau)^2}}+\frac{\gamma_1 A(\tau)^3}{\sqrt{2/w^2+8 A(\tau)^2}} e^{\frac{-2k_L^2 s_0}{1+4w^2 A(\tau)^2}}\right).
\end{eqnarray}
For homogeneous nonlinear interaction ($\gamma_1=0$) $V_{\rm eff}$ in  $(9)$ gives a periodic potential. The spatial extension of this potentail depends on $k_L$. On the other hand, the presence of spatially inhomogeneous interaction ($\gamma_1\neq 0)$ gives rise to an effective potential (second term in $V_{\rm eff}$) the effect of which is likely to depend on the spatial width ($w$) of the (Gaussian) nonlinearity. When the spatial inhomogeneous atomic interaction acts in a BEC loaded in an optical lattice, then such an interaction will tend to  change both the shape and magnitude of the trapping potential arising from the optical lattice. Therefore, the relative values of $k_L$ and $w$ may be expected to  have considerable effects on $V_{\rm eff}$ and thus on the dynamics of the solitons.
\par
With a view to get a physical feeling from $V_{\rm eff}$ on the behaviour of the soliton due to interplay between the optical lattice and spatial nonlinearity, we consider two cases, namely,  (i) $w>k_L$ and (ii) $w<k_L$. In both $(i)$ and $(ii)$ we shall focus on whether $\lambda>A^{-1}$(= soliton's width) or $\lambda(=2\pi/k_L)<A^{-1}$.
\begin{figure}[!htb]
\begin{center}
\includegraphics[width=10cm]{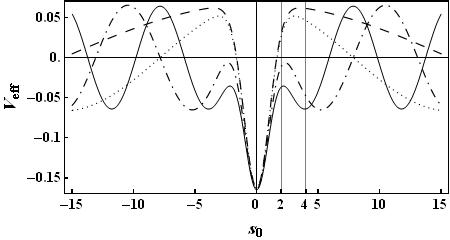}
\caption{\small Effective potential resulting from OL and localized inhomogeneous interaction with $A=2$, $V_0=0.2$,$\gamma_1=-0.5$, $w=1$ and for $k_L=$ $0.05$(dashed), $0.1$(dotted), $0.3$(dot-dashed) and $0.4$(solid).}
\end{center}
\end{figure}
For the first case, the variation of $V_{\rm eff}$ as a function of $s_0$
for different $k_L$ values corresponding to $\lambda>A^{-1}$ is displayed in figure $1$. In this figure, the dashed, dotted, dot-dashed and solid curves show  $V_{\rm eff}$  with  $V_0=0.2$, $w=1$ and $\gamma_1=-0.5$ for $k_L=0.05,\, 0.1,\,0.3$ and $0.4$ respectively. The variational approach followed by us shows that the amplitude of the soliton remains constant with time. In view of this, we shall work with a constant values of $A$, namely, $2$. From figure $1$ we see that each curve  for the effective potential consists of two distinct parts. A deep potential well around $s_0=0$ and a periodic potential with smaller amplitudes on either sides of the well. Clearly, periodicity of the effective potential increases as $k_L$ increases. With the increase of $k_L$, $V_{eff}$ begins to deform near the  well  due to interplay between the OL and spatially localized atomic interaction. Particularly, for $k_L=0.1$ there appears two peaks (dotted line), the amplitudes of which depend on $k_L$. As $k_L$ increases amplitudes of these peaks decrease (dot-dashed line). For $k_L=0.4$, amplitude of these peak (solid curve) becomes too low that a soliton with sufficient  energy initially situated at the minima near the well may cross the peak  and will begin to move inside the well.
\par
On the other hand, the variation of $V_{\rm eff}$ as a function of $s_0$ with $A=0.5$ for different $k_L$ values corresponding to $\lambda<A^{-1}$ displayed in figure $2$.
\begin{figure}[!htb]
\begin{center}
\includegraphics[width=8.5cm]{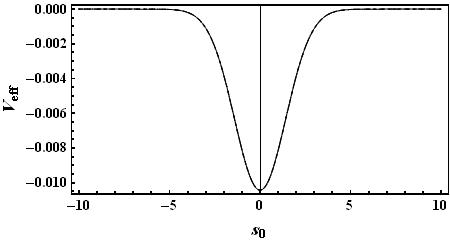}
\caption{\small Effective potential resulting from OL and localized inhomogeneous interaction with $A=0.5$, $V_0=0.2$,$\gamma_1=-0.5$, $w=1$ and for different values of $k_L$, namely, $0.05$, $0.1$, $0.3$ and $0.4$. Clearly, all curves merge and form a single curve.}
\end{center}
\end{figure}
This figure shows that the curves for different $k_L$ coincide and appear to form a single curve with only a deep well around $s_0=0$. Clearly, the effects of the OL is almost negligible. As a result, we have only a potential well produced due to the localized nonlinear interaction. Since our aim in this paper is to study combined effects of the OL and spatially localized atomic interaction, we shall have to work with $\lambda<A^{-1}$. Note that the potential curve in figure $2$ has a minimum at $s_0=0$. So, one may expect that the soliton in this case  may also be stable there. We have verified that the optical trap is unable to confine such solitons and thus they are unstable.
\begin{figure}[!htb]
\begin{center}
\includegraphics[width=8.5cm]{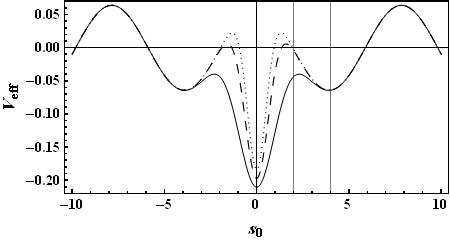}
\caption{\small Effective potential resulting from OL and localized inhomogeneous interaction with $A=2$, $V_0=0.2$,$\gamma_1=-0.5$, $k_L=0.4$ and for  $w=$ $0.5$(dotted), $0.6$ (dashed) and $1$ (solid).}
\end{center}
\end{figure}
\par
In view of the above, for $w>k_L$ we consider only the case $\lambda>A^{-1}$. In this case, the plot of $V_{\rm eff}$ as a function of $s_0$  for different values of $w$  is shown in figure $3$. Here, values of the other parameters  are kept same as those used in figure $1$ but $k_L=0.4$.  In this figure, the dotted, dashed and solid curves give  variations of $V_{\rm eff}$ for $w=0.5$, $0.6$ and $1$ respectively. Clearly, all the curves  represent periodic potentials with local minima at $s_0=0$. These local minima are separated by potential barriers from the minima of lattice potential. Interestingly, the depth of the intermediate (the central minimum) well introduced by the inhomogeneous interaction increases as $w$ increases.  But as $w$ increases the  heights $(h)$ of these barriers decrease. At $w=1$, the value of $h$ becomes so small that the soliton with  sufficient initial energy can cross the barrier and enter into the well. Therefore, it is clear that the crossover from a lattice minimum to the interaction energy minimum can also be taken place with the variation of laser-beam width.
\par
The  potential curves we have portrayed above are the effective potentials for the soliton in  a single-component BEC where all the atoms are in the same hyperfine state. However, it has also been possible to confine atoms of different hyperfine states or different species in a single trap leading to  a two-component BEC \cite{30}. In addition to inter-atomic interactions in a single-component BEC, the interaction between the atoms of different component (inter-species) in the two-component BEC  has come into play. As a result, the effective potential for a component soliton
in a coupled BEC will be affected due to this inter-species interaction. It is of interest to note that the effective potential in figure $1$ has some similarity with the effective potential for the separation between of centers  of the solitons in a two-component BEC with optical lattices \cite{13,31}. For example, the $V_{\rm eff}$ in both cases consists of two parts, (i) a deep potential well and (ii) oscillatory wings on either side of the well. Interestingly,  source of the periodic part of the effective potentials for two interacting solitons as well as for the soliton in the present case is same. However, origin the well in effective potentail in two cases are different. In the present case, this  well (figure $1$) appears due to presence of spatially localized inter-atomic interaction. On the other hand, it is shown in refs. $12$ and $33$ that the  same well in the effective potential arises due to mutual interaction between the condensates. Therefore, the mutual interaction between the solitons in a coupled condensates may reasonably be  approximated by a spatially localized interaction. Whatever the origin of the well in $V_{\rm eff}$, we see that the response of solitons to the variation of soliton and lattice parameters in both cases is almost same. For example, the solitons in both cases can be forced to move in localized/mutual interaction region only by varying the wave number $k_L$. A soliton of a relatively larger width appears to experience an effective potential (figures $2$, a potential having no oscillatory wings) similar to the  potential that arises  due to only mutual interaction between the solitons in a coupled BEC. Therefore, the system considered here  may server to a certain extend as an useful model to get some realization on the behaviour of coupled solitons loaded in (linear or nonlinear) optical lattices. However, to confirm this remark we shall have to study dynamics of the soliton for different environments of the effective potential.
\section{Dynamics of matter-wave bright solitons}
We have seen that the bright soliton in a BEC with spatially localized nonlinearity, loaded in OLs  experiences  an effective potential $V_{\rm eff}$ which consists of two parts, (i) a deep well around $s_0=0$ and (ii) oscillatory wings on either side of the well. Clearly, the well arises due to the inhomogeneous nonlinearity while the oscillation in $V_{\rm eff}$ due to the OL. We also pointed out that the variation of wave number $k_L$ of the OL as well as width $w$ of the Gaussian nonlinearity deforms $V_{\rm eff}$ considerably only near the well at $s_0=0$. Therefore, it will be an interesting curiosity  to investigate how the soliton  responses to such deformed periodic potential. To visualize this, we solve the ordinary differential equation in $(8b)$ using Runge-Kutta method for different  values of $s_0$ and ${\dot{s}}_0$ at $\tau=0$ and calculate density profiles of the solitons. We also  solve GP equation in $(1)$ using a purely numerical routine to verify the result obtained from variational calculation. In particular, we shall make use of Crank-Nicholson method to solve $(1)$ with the conditions $\phi(s,0)=sech(s-s_0) exp[i\dot{s}_0(s-s_0)]$, $\phi(-25,\tau)=\phi(25,\tau)=0$  and $\gamma_0=-1$. Note that the numerical calculation of  $(1)$ not only serves as a supplement of the result obtained  on the basis of  a variational approach but also explicitly show whether the density profile will decay or not with time parameter $\tau$.
\par
The effective potentials ($V_{\rm eff}$) shown in figures $(1)$ and $(3)$  are equivalent in the sence of their shapes. The only exception is that in the first case the deformation in $V_{\rm eff}$ arises with $k_L$ while in the later it is due to variation of $w$. In view of this, we consider only figure $1$ (solid curve) to calculate different data required in calculating density profiles $|\psi|^2$. The influence of the inhomogeneous interaction on lattice potentials is prominent near $s_0=0$. In order to visualize the effects arising due to interplay between the nonlinearity and lattice on the condensate, we shall focus our attention on the minimum nearest of the central minimum at $s_0=0$.  In figure $4$, we display $|\phi(s,\tau)|^2$ for an initial position of the soliton center at the local minimum of $V_{\rm eff}$ around $s_0(0)=3.9$ having initial velocity ${\dot{s}}_0(0)=0$. In the same figure we also display variations of the soliton's center and velocity with $\tau$. In particular, figure $4(a)$ shows the plots of $s_0$ and ${\dot{s}}_0$ as a function of $\tau$ while  figures $4(b)$ and $4(c)$ display the evolution of density profiles $|\phi(s,\tau)|^2$ obtain from numerical and variational calculations. Figure $4(a)$ clearly shows that position and velocity of the soliton remain unaltered with time $\tau$. On the other hand, the variation of density in figures $4(b)$ and $4(c)$ indicate that the solitonic evolution is not affected by the lattice periodicity as well as inhomogeneous interaction. Note that analytical and numerical results are identical. The   evolution of density profiles with unchanged amplitude for a larger scaled time $\tau$ manifest that the solitons are stable. We have verified that this behavior of  solitons is same for all other local minima of the potential.
\begin{figure}[!htb]
\begin{center}
\includegraphics[width=5.5cm]{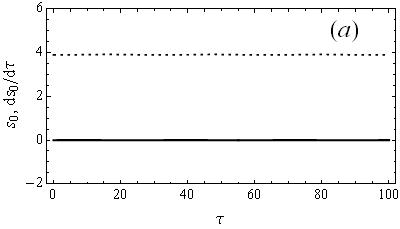}{\hspace*{0.5cm}}
\includegraphics[width=5cm]{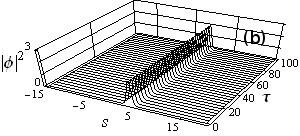}{\hspace*{0.5cm}}
\includegraphics[width=5cm]{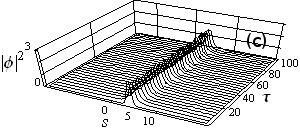}
\caption{\small $(a)$ Variations of position (dotted curve) and velocity (solid curve) of a soltion with $\tau$ for $s_0(0)=3.9$ and ${\dot{s}}_0(0)=0$.  Density profiles $|\phi(s,\tau)|^2$ of solitons with  $\tau$ obtained from $(b)$ variational and $(c)$ numerical calculations.}
\end{center}
\end{figure}
\par
Now let us see what happens if the soliton having $s_0(0)=3.9$ possesses very small initial velocity, say  ${\dot{s}}_0(0)=0.4$, instead of zero. In this case, variations of position and velocity of the soliton is shown in figure $6(a)$ while figures $6(b)$ and $6(c)$ display evolution of density profiles with $\tau$ obtained respectively from variational and numerical approaches.
\begin{figure}[!htb]
\begin{center}
\includegraphics[width=5.5cm]{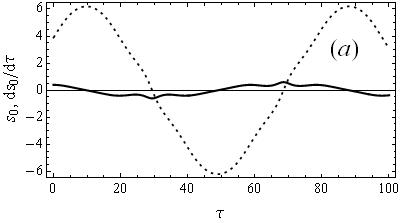}{\hspace*{0.5cm}}
\includegraphics[width=5cm]{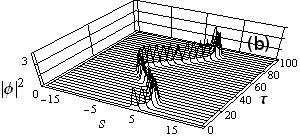}{\hspace*{0.5cm}}
\includegraphics[width=5cm]{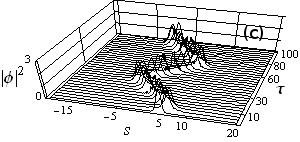}
\caption{\small $(a)$ Variations of position (dotted curve) and velocity (solid curve) of a soltion with time for $s_0(0)=3.9$ and ${\dot{s}}_0(0)=0.1$. Density profiles $|\phi(s,\tau)|^2$ of solitons with  $\tau$ obtained from $(b)$ variational and $(c)$ numerical calculations.}
\end{center}
\end{figure}
From  $5(a)$ we see that the position and velocity of the soliton  rather than being constant start to vary. Clearly, the soliton first crosses the peak of the $V_{\rm eff}$ (figure $1$) and moves toward $s_0=0$ with increasing velocity. At $s_0=0$, its velocity becomes maximum. From this point, the soliton again starts to move up the potential slope with decreasing velocity but in the opposite direction. The motion  continues until the soliton gets the initial velocity. This process goes on repeatedly. As a result, the soliton oscillates around the minimum of the potential well. The plots of  $|\phi(s,\tau)|^2$ in figures $5(a)$ and $5(b)$ show that variational calculation again agree with the numerical results. This figure also indicates that during  evolution the soliton remains stable for a long time.
\begin{figure}[!htb]
\begin{center}
\includegraphics[width=5.5cm]{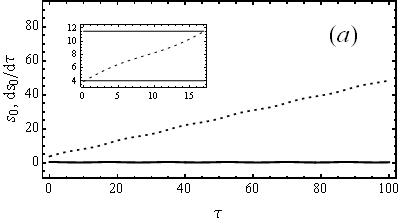}{\hspace*{0.5cm}}
\includegraphics[width=5cm]{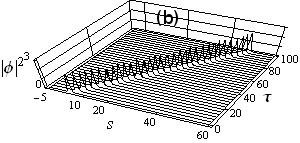}{\hspace*{0.5cm}}
\includegraphics[width=5cm]{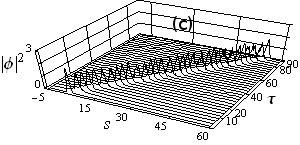}
\caption{\small $(a)$ Variation of position (dotted curve) and velocity (solid curve) of a soltion  with time for $s_0(0)=3.9$ and ${\dot{s}}_0(0)=1$.  Density profiles $|\phi(s,\tau)|^2$ of solitons with  $\tau$ obtained from $(b)$ variational and $(c)$ numerical calculations.}
\end{center}
\end{figure}
\par
It may be quite interesting to investigate the behaviour of the soliton if its initial velocity is relatively large, say ${\dot{s}}_0(0)=1$ having same initial position. This behaviour is displayed  in figure $6$. Here $6(a)$ shows variation of $s_0$ and ${\dot{s}}_0$ with $\tau$ while $6(b)$ and  $6(c)$ give evolution of density profiles of the solitons. Clearly, for such initial velocity the soliton instead of moving towards the potential well at $s_0=0$, it moves away from $s_0=3.9$ with almost a constant speed. As a result, $s_0$ only increases as $\tau$ increases. However, while propagating the soliton shows oscillatory motion in space having periodicity  same as that of the lattice potential. The numerical calculation (figure $6(c)$) here also confirms the results in figure $6(b)$. The unattenuated propagation of the density profiles imply that solitons are stable.
\begin{figure}[!htb]
\begin{center}
\includegraphics[width=5.5cm]{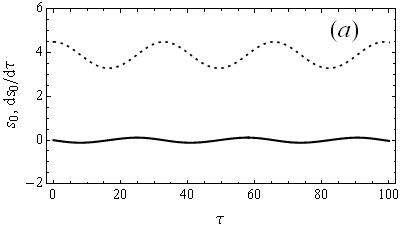}{\hspace*{0.5cm}}
\includegraphics[width=5cm]{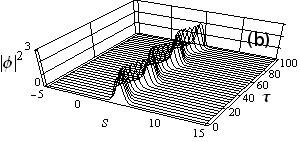}{\hspace*{0.5cm}}
\includegraphics[width=5cm]{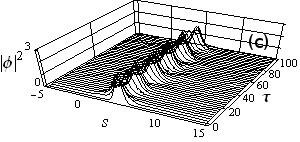}
\caption{\small $(a)$ Variation of position (dotted curve) and velocity (solid curve) of a soltion with time for $s_0(0)=4.5$ and ${\dot{s}}_0(0)=0$.
Density profiles $|\phi(s,\tau)|^2$ of solitons with  $\tau$ obtained from $(b)$ variational and $(c)$ numerical calculations.}
\end{center}
\end{figure}
\par
So far we have studied the dynamics of solitons with respect to a minimum near the  well centred at $s_0=0$ of the effective potential. However, it remains to test the behaviour of the soliton for  initial locations other than the minimum of effective potential.  We plot in figure $7(a)$ variations of  $s_0$ and ${\dot{s}}_0$ for ${\dot{s}}_0(0)=0$ and $s_0(0)=4.5$ while  $7(b)$  and $7(c)$  give density profiles $|\psi(s,\tau)|^2$  versus $\tau$. Figure $7(a)$ clearly shows that both $s_0$ and ${\dot{s}}_0$ vary periodically with $\tau$. However, unlike  the previous case the periodicity of such motion depends on the displacement of the soliton from its nearest local potential minimum ($s_0=3.9$). In particular, they exhibit oscillatory motion about the nearest local potential minimum. It may be noted that the behaviour of the soliton remains same for any values of $s_0$ which do  not correspond to potential extrema.
As in the previous case, the density profiles obtained from variational [$7(b)$] and numerical [$7(c)$] calculations are identical and exhibit stable oscillatory motion.
\section{Gap soliton solution}
An important feature of a BEC in OLs is that it can hold a novel type of spatially localized state which is bright-soliton like and can exist even in a repulsive BEC \cite{32}. This soliton resides inside a gap of the band-gap spectrum of linear version of the GP equation with a periodic potential and thus they are called gap solitons (GS). Understandably, this type of localized state exits due to balance between self-repulsion and negative effective mass provided by the OL spectrum and the chemical potential of GS must fall in the spectrum's band gap, where delocalized Bloch-wave states can not exist. The creation of gap soliton in Q1D BEC has also been reported in \cite{15,33}. Here, the BEC was pushed in a state with negative effective mass by means of acceleration.
\par
In this section, we shall consider fundamental gap solitons in a Q1D BEC with spatially localized interaction. The fundamental gap solitons \cite{34} are indeed compact objects with quasi-Gaussian shapes, trapped in a single cell of the optical lattice potential. Such gap solitons in the presence of spatially localized interaction can be descrided by means of the GP equation $(1)$. The stationary version of $(1)$ is given by
\begin{equation}
\mu \psi=-\frac{1}{2}\frac{\partial^2 \psi}{\partial s^2}+ V_0 \cos(2 k_L s)\psi +\left( \gamma_0+\gamma_1 exp\left[-s^2/2 w^2\right]\right)|\psi|^2 \psi.
\end{equation}
Here, $\mu$ is the chemical potential. In writing $(10)$ we have used $\phi(s,\tau)=\psi(s) e^{-i\mu\tau}$ in $(1)$ with $\int_{-\infty}^{+\infty} |\psi(s)|^2 ds=N$. To proceed with the variational approach we first assign a Lagrangian of $(10)$. One can cheek that
 the Lagrangian
\begin{equation}
L=\mu \psi^2 +\frac{1}{2} \left(\frac{d\psi}{d\tau}\right) ^2+V_0 \cos(2 k_L s)\psi^2+\frac{1}{2}\left(\gamma_0+\gamma_1 exp\left[-s^2/2 w^2\right]\right)\psi^4
\end{equation}
via the Euler-Lagrange equation gives $(10)$. In order to find a trial solution of $(10)$, note that the fundamental gap soliton we considered is, in fact, have the form of a bright soliton. Therefore, the trial solution for the gap soliton can also be adopted as
\begin{equation}
\psi(s)=\sqrt{\frac{N}{\sqrt{\pi} a}} e^{-s^2/{2 a^2}},
\end{equation}
where variational parameters $N$ and $a$ are respectively the amplitude and width of the soliton. Integrating the equation obtained by inserting $(12)$ in $(11)$ over $s$ from $-\infty$ to $+\infty$ we get the averaged Lagrangian $\left\langle {L}\right\rangle$ in the form
\begin{eqnarray}
\left\langle {L}\right\rangle =N\left(-\mu+\frac{1}{4 a^2}+V_0
 e^{-k^2_L a^2} +
 \frac{\gamma_0 N}{2 a\sqrt{2\pi}}+
\frac{\gamma_1 N w}{a\sqrt{2\pi}\sqrt{ 4 w^2+a^2}}\right).
\end{eqnarray}
From the vanishing conditions of $\frac{d\left\langle {L}\right\rangle}{dN}$ and $\frac{d\left\langle {L}\right\rangle}{da}$  we get
\begin{equation}
\mu=\frac{1}{4 a^2}+ V_0 e^{-a^2 k_L^2}+
\frac{\gamma_0 N}{2 a\sqrt{2\pi}}+\frac{\sqrt{2}\gamma_1 N w}{a\sqrt{\pi}\sqrt{ 4 w^2+a^2}}
\end{equation}
and
\begin{equation}
V_0=\frac{e^{a^2 k_L^2}}{8 a k_L^2}\left( \frac{2}{a^3}+\sqrt{\frac{2}{\pi}}\frac{\gamma_0 N}{a^2}+\sqrt{\frac{2}{\pi}}\frac{4\gamma_1 N w(a^2+2w^2)}{a^2 (a^2+4 w^2)^{3/2}}\right).
\end{equation}
For $\gamma_0=\gamma_1=0$, $(14)$ and $(15)$ give
\begin{eqnarray}
V_0=\frac{1}{4 a^4 k_L^2} e^{-a^2k_L^2}
\end{eqnarray}
and
\begin{eqnarray}
\mu=\frac{1}{4 a^2}+ V_0 e^{-a^2 k_L^2}.
\end{eqnarray}
Now, we make use of $(16)$ and $(17)$ to draw the band gap spectrum. The band-gap diagram for $k_L=1$ is displayed in figure $8$.
\begin{figure}[!htb]
\begin{center}
\includegraphics[width=8cm]{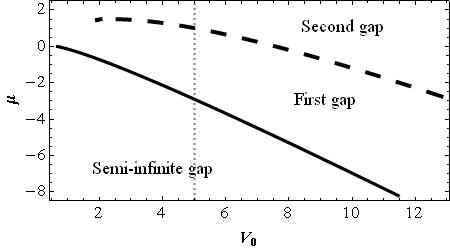}
\caption{\small Band - gap spectrum of the linearized GP equation in $(10)$. In this diagram we replaced $V_0$ by $-V_0$. The solid line between semi-infinite and first gap is obtained for even soliton while the dotted line separating first and second gap is drawn by considering odd solitons solution \cite{34}.}
\end{center}
\end{figure}
Clearly, variational method gives the well-known band-gap spectrum for the  linearized version of $(10)$. In order to obtain fundamental gap soliton solution we take $V_0$ and $\mu$ values in the first band gap and solve $(10)$  with the initial conditions $\psi(s)|_{s=0}=1$ and  $\frac{d\psi}{ds}|_{s=0}=0$. Here, we have used $V_0=-5$ and $\mu =-2$. Clearly, the value of $\mu$ lies in the first gap (figure $8$). The result obtained form the numerical integration of $(1)$ is shown in figure $9$. We are interested to fundamental gap solitons in a BEC with repulsive atomic interaction i.e, with positive $\gamma$ values. In view of this, we plot in figure $9(a)$ $|\psi(s)|^2$ as a function of $s$ for $\gamma_0=0.77$ and  $\gamma_1=0.5$.
\begin{figure}[!htb]
\begin{center}
\includegraphics[width=7.5cm]{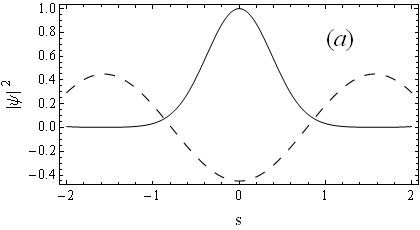}{\hspace*{1cm}}
\includegraphics[width=7.5cm]{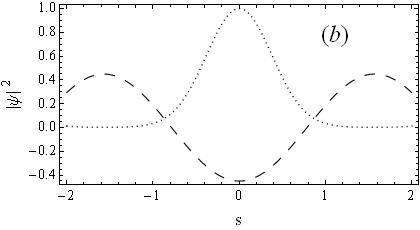}
\caption{\small Profiles of fundamental gap solitons for $w=1$, $k_L=1$ with strength of repulsive atomic interaction $(a)$ $\gamma_0=0.77$ and $\gamma_1=0.5$ (solid curve) and  $(b)$ $\gamma_0=0 $ and $\gamma_1=1.295$ (dotted curve)}
\end{center}
\end{figure}
Here, the solid curve represents  $|\psi(s)|^2$ while the  dashed curve serves as a guide of the lattice potential. We have noted that the spatially localized atomic interaction have some effects on the effective potential for the bright solitons. With a view to examine effects of the localized interaction on the fundamental gap soliton we make $\gamma_0=0$, arising background scattering length $a_0$. In this case, the density profile obtained from numerical integration of $(10)$ is portrayed in figure $9(b)$.
The dotted curve in  figure $9(b)$ clearly  shows that the density profile here is also localized. Interestingly, such localized wave exists only for relatively higher  $\gamma_1$ values, namely, $1.295$.
Therefore, a BEC in OLs with spatially localized nonlinear interaction may support quasi-Gaussian shaped solution.
\par
Up to now it is not clear whether the fundamental gap solitons discussed above  are stable or not. In order to examine stability of such localized states we make use of Vakhitov-Kolokolov criterion \cite{35}. According to this  criterion  localized nonlinear wave like soliton will be stable if  $\partial \mu/\partial N<0$. However,  studies on fundamental gap soliton reveal that such solitons are linearly stable as long as $\partial \mu/\partial N>0$ \cite{36}.  Understandably, the generalized Vakhitov-Kolokolov (VK) criterion is ``inverted'' for a stable fundamental gap soliton. This  inverted VK criterion has also been explained by the approximation based on the averaging method \cite{20}. Here we shall exploit variational method to justify stability of the profiles portrayed in figures $9(a)$ and $9(b)$ and also to verify inverted VK criterion for the gap soliton considered.
\par
From $(14)$ and $(15)$ we can obtain
\begin{equation}
\frac{d\mu}{d N}=\frac{\gamma_0}{a \sqrt{2\pi}}(1-\frac{1}{4 a^2 k_L^2})+\sqrt{\frac{2}{\pi}}\frac{\gamma_1 w}{a\sqrt{a^2+4 w^2}}\left( 1-\frac{(1+a^2)}{4k_L^2a^2(a^2+4 w^2)}\right).
\end{equation}
In figure $(10)$ we plot $d\mu/dN$ as a function of $a$ for $w=1$ and $k_L=1$. In particular, the solid curve shows $d\mu/dN$ in presence of both  homogeneous and inhomogeneous atomic interactions ($\gamma_0\neq0$ and $\gamma_1\neq0$) while the dotted curve gives similar plot for $\gamma_0=0$ but $\gamma_1\neq0$.
\begin{figure}[!htb]
\begin{center}
\includegraphics[width=8.5cm]{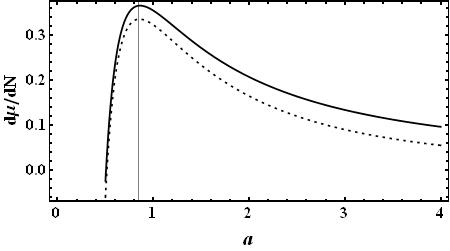}
\caption{\small Variation of $d\mu/dN$ as a function of $a$ for (i) $\gamma_0=0.77$ and $\gamma_1=0.5$ (solid curve) and (ii) $\gamma_0=0$ and $\gamma_1=1.295$ (dotted curve)}
\end{center}
\end{figure}
Both the curves in this figure  clearly show $d\mu/dN$ has a positive maximum value at $a=1$ (indicated by vertical line). On either sides of $a=1$, the value of $d\mu/dN$ gradually decreases. Sivan et al \cite{10} showed that the solitons are more stable at the maximum value of $d\mu/dN$. Therefore, the fundamental gap solitons considered here are stable if their width $a\approx 1$. From the figures $9(a)$ and $9(b)$, we see that the width of  density profiles of the fundamental gap solitons obtained from direct numerical calculations have width $a\approx1$ and thus they are stable. Understandably, the positive value of $d\mu/dN$ implies that the stable fundamental gap soliton in the presence of spatially localized interaction also satisfy ``inverted'' Vakhitov-Kolokolov criterion. Recently, a physical mechanism of scattering of the gap solitons in the presence of linear Gaussian impurity has been reported in \cite{37}. In particular, they observe the occurrence of repeated reflection, transmission, and trapping regions due to resonance of the linear impurity mode with gap solitons. Our study on the existence of gas soliton in the presence of nonlinear Gaussian impurity may motivate for further study in this direction.
\section{Conclusion}
We derived an expression of the effective potential for a soliton in  BECs which were confined in an OL having spatially localized atomic interactions with a view to  study the dynamics of the soliton. The localized interaction introduces a deep well at a maximum of the periodic potential due to the optical lattice. This well is separated from the nearest lattice minima by potential wall. The amplitude of which depends on lattice parameter (namely, wave number) as well as width of localized inter-atomic interaction. Therefore, if a soliton initially located at a minimum nearest to the well is at rest, it can be forced to move into the well by varying only the width of localized interaction.
\par
We showed that dynamical behavior of the soliton depends sensitively on its initial location in effective potential as well as its velocity. This is obtained by  solving the appropriate GP equation by a purely numerical routine. We examined that, depending on the initial conditions,  the density profile of the soltion may execute steady and oscillatory motions and can move with a constant speed without any attenuation. The unattenuated long time evolution of the density profile implied that during evolution the profile remains stable. In addition, the numerical calculations not only gave validity of the method used but also justified the accuracy of the trial function for the model considered.
\par
We also showed that the effective potential resulting from the localized nonlinearity and OL for a soliton in a single-component BEC is almost same as that of the effective potential of two interacting solitons in a two-component BEC with optical lattices. Interestingly, we found that the response of the soliton was also similar to that of the solitons in  two-component BECs. Therefore, the effective potential for two interacting solitons in a BEC may be represented by a Gaussian space-dependent function.
\par
Finally, we noticed that the system can support fundamental gap solitons even in the absence of background scattering length and/or even only in the presence of spatially localized interaction. These solitons are stable only when they satisfy inverted Vakhitov-Kolokolov criterion.

\end{document}